\documentstyle[12pt,aaspp4]{article}
\newcommand\beq{\begin{equation}}
\newcommand\eeq{\end{equation}}

\begin{document}

\title{Bondi Accretion and the Problem of the Missing Isolated Neutron Stars}

\author{Rosalba Perna\altaffilmark{1,2}, Ramesh Narayan\altaffilmark{2}, 
George Rybicki\altaffilmark{2}, Luigi Stella\altaffilmark{3} and Aldo Treves\altaffilmark{4}}

\altaffiltext{1}{Harvard Society of Fellows, 74 Mount Auburn Street, Cambridge, MA 02138}
\altaffiltext{2}{Harvard-Smithsonian Center for Astrophysics, 60 Garden Street,
Cambridge, MA 02138}
\altaffiltext{3} {Osservatorio Astronomico di Roma, Sede di Monteporzio Catone, 
Via Frascati 33, I-00040 Rome, Italy}
\altaffiltext{4}{Dipartimento di Scienze, Universita' dell'Insubria, Via Valleggio 
11, I-22100 Como, Italy}

\begin{abstract}
A large number of neutron stars (NSs), $\sim 10^9$, populate the
Galaxy, but only a tiny fraction of them is observable during the
short radio pulsar lifetime.  The majority of these isolated NSs, too
cold to be detectable by their own thermal emission, should be visible
in X-rays as a result of accretion from the interstellar medium.  The
ROSAT all sky survey has however shown that such accreting isolated
NSs are very elusive: only a few tentative candidates have been
identified, contrary to theoretical predictions that up to several
thousands should be seen. We suggest that the fundamental reason for
this discrepancy lies in the use of the standard Bondi formula to
estimate the accretion rates.  We compute the expected source counts
using updated estimates of the pulsar velocity distribution, realistic
hydrogen atmosphere spectra, and a modified expression for the Bondi
accretion rate as suggested by recent MHD simulations, and supported
by direct observations in the case of accretion around supermassive
black holes in nearby galaxies and in our own.  We find that, whereas
the inclusion of atmospheric spectra partly compensates for the
reduction in the counts due to the higher mean velocities of the new
distribution, the modified Bondi formula dramatically suppresses the
source counts.  The new predictions are consistent with a null
detection at the ROSAT sensitivity.

\end{abstract}

\section{Introduction}

Neutron stars (NSs) are a common endpoint of stellar evolution.
Nucleosynthesis constraints on Galactic chemical evolution require
that the total number of NSs in our Galaxy today be on the order of
$10^9$ (Arnett, Schramm \& Truran 1989).  After they are born, NSs are
visible for a short while either as X-ray sources from their cooling
radiation or as radio sources from pulsar activity.  These phases,
however, only last $\sim 10^6$ and $\sim 10^7$ yr, respectively, which
is a small fraction of the age $\sim 10^{10}$ yr of the Galaxy.  After
a few tens of millions of years both the thermal and the rotational
energy supply will be exhausted, and from then on the NSs would just
be cold, invisible objects, if it were not for accretion from the
interstellar medium (ISM), which makes them shine again in the X rays,
as it was realized some 30 years ago by Ostriker, Rees \& Silk (1970).
The problem received close attention in the early 1990s. According to
the calculations of Treves \& Colpi (1991) and Blaes \& Madau (1993;
BM in the following), the luminosity due to accretion can be
significant, and as many as  10,000 NSs should be
detectable with the ROSAT all sky survey.

The problem of the ``Missing Neutron Stars'' became evident when the
observations returned only a handful of possible candidates.
Subsequent revisions of the expected rates have somewhat alleviated
the problem by accounting for other effects not included in the BM
calculations, such as dynamical heating (Madau \& Blaes 1994),
coupling of the dynamical and magnetorotational evolution (Popov et
al. 2000; Livio, Xu \& Frank 1998; Colpi et al. 1998).  
Furthermore, in the last few years, more accurate
estimates of pulsar velocities (Lorimer, Bailes \& Harrison 1997;
Hansen \& Phinney 1997; Cordes \& Chernoff 1998; Arzoumanian et al. 2002) seemed to point
towards an increase in the mean velocity (and hence a decrease in the
accretion rate) relative to the Narayan \& Ostriker (1990)
distribution used by BM.  Whereas larger typical velocities imply
smaller accretion luminosities, on the other hand, as we show here
(see also Zane et al. 1995), the inclusion of more realistic spectra
which go beyond the simple blackbody spectrum, tends to increase the
predicted X-ray number counts\footnote{This is due to the fact that,
for the same total luminosity, the spectrum from a light element
atmosphere is significantly harder than a blackbody (see \S 2.3 for
more details).}  and hence partly counterbalance the reduction due to
other effects.

In this paper, we propose that a likely solution to the missing
neutron stars problem lies in the use of the Bondi expression to
calculate the accretion rate.  Since the work of Fabian \& Canizares
(1988), it has been known that the accretion luminosities of
supermassive black holes in most galactic nuclei are much too low to
be consistent with radiatively efficient accretion at the Bondi rate
from the surrounding ISM.  The evidence has become compelling in
recent years (Di Matteo et al. 2000, 2001; Loewenstein et
al. 2001; Narayan 2002).  Two solutions to this problem have been
proposed (see e.g. Narayan 2002 for a review): (i) Radiatively
inefficient accretion, as in an advection-dominated accretion flow
(Ichimaru 1977; Rees et al. 1982; 
Narayan, Yi \& Mahadevan 1995; Fabian \& Rees 1995); (ii) Accretion
at below the Bondi mass accretion rate (Blandford \& Begelman 1999;
Narayan, Igumenshchev \& Abramowicz 2000; Quataert \& Gruzinov 2000).
The current consensus is that both effects operate, with radiatively
inefficient accretion triggering a reduction in the mass accretion
rate $\dot M$.  The case for a reduced $\dot M$ has become especially
strong with the recent detection of linear polarization at 230 GHz
from Sgr A$^*$ at the Galactic Center (Bower et al. 2003).  This
observation indicates that the gas density close to the black hole is
much less than that predicted for Bondi accretion.

In the context of the missing neutron star problem, we note that
radiatively inefficient accretion alone will not solve the problem
since the advected energy will be radiated once the accreting gas hits
the surface of the NS.  A reduction in the mass accretion rate
relative to the Bondi rate is essential.  Such a reduction has been
seen in recent MHD simulations of spherically symmetric accretion with
weak, dynamically unimportant
magnetic fields (Igumenshchev \& Narayan 2002).  The radial structure
of the accretion flow is dramatically modified from the Bondi profile,
and the mass accretion rate is decreased significantly.  Other work on
rotating flows also gives similar results (\S2.3 below).

Backed by the above observational and theoretical results, we
perform a new calculation of the count rates for isolated NSs that
accounts for the recent estimate of the velocity distribution, more
realistic atmospheric spectra, as well as the correction to the Bondi
rates predicted by the numerical simulations.  We find that the
resulting numbers are consistent with null detection at the ROSAT
sensitivity.

\section{Dynamical evolution, emission properties and accretion rates}

The computation of the number counts requires several ingredients, and
we describe each of them in detail in the following.

\subsection{Dynamical evolution and local space density of INSs}

As discussed in \S 1, in the last few years several groups (Lorimer,
Bailes \& Harrison 1997; Hansen \& Phinney 1997; Cordes \& Chernoff
1998; Arzoumanian et al. 2002) have realized that the velocity
distribution for pulsar velocities derived by Narayan \& Ostriker
(1990; NO in the following) overpredicts the fraction of slow moving
stars and underpredicts the contribution from faster moving
ones. Here, we adopt the estimates by Cordes \& Chernoff (1998, CC in
the following).  Using a sample of 49 young pulsars, they found that
the distribution $p(v_{Ri},v_{\phi i},v_{zi})$ for the velocities of
their progenitors can be best described by a two-component gaussian
model with characteristic speeds of $\sim 175$ km/s and $\sim 700$
km/s, representing $\sim 86\%$ and $14\%$ of the population,
respectively.  The scale height for both components (assuming a
gaussian distribution) is $\sim 0.13$ kpc.  The radial birth
distribution is assumed to follow that of the stellar disk, that is an
exponential with scale length 4.5 kpc out to a maximum radius of 20
kpc.

For a given initial phase space distribution
$f_i(R_i,z_i,v_{Ri},v_{\phi i}, v_{zi})$, the equilibrium distribution
can be determined by Monte Carlo simulations of the orbits of the
stars in the galactic potential (e.g. Paczynski 1990; Blaes \&
Rajagopal 1991).  However, this is a very time-consuming method and,
as BM demonstrated, an excellent approximation to the numerical
simulations is provided by the use of Liouville's theorem under the
``thin-disk approximation'', where the motions along $R$ and $z$ can
be decoupled. In this case, the equilibrium phase space distribution
is related to the initial one through the relation \beq \int
dR\,dz\,dv_{R}\,dv_\phi\,dv_z\,f_{\rm eq}(L_z,E_R,E_z)= \int
dR_i\,dz_i\,dv_{R i}\,dv_{\phi i}\,dv_{z
i}\,f_i(R_i,z_i,v_{Ri},v_{\phi i},v_{zi})\;,
\label{eq:dist}
\eeq
where the integrals are taken over the regions of phase space around
given values of $L_z$, $E_R$, and $E_z$.

Figure 1 shows a comparison between the equilibrium speed distribution
derived from the NO initial distribution, and the one derived from the
CC birth distribution. It can be seen that there is a significantly
larger fraction of high velocity stars.  When using the birth
parameters derived by CC, the local NS density $n_{\rm
NS}(R=8.5\;{\rm kpc},z=0)\equiv n^0_{\rm NS}$ is reduced from the value
$n^0_{\rm NS}\approx 7\times 10^{-13} N_{\rm tot}$ ${\rm pc}^{-3}$ found by BM for the NO
population, to $n^0_{\rm NS}\approx 4\times 10^{-13} N_{\rm tot}$ ${\rm pc}^{-3}$.

It should be noted that the relation in equation (\ref{eq:dist}) does
not account for the fraction of high velocity stars that escape the
Galaxy. As a matter of fact, this has very little effect on the
statistics of the high-$\dot{M}$ tail of the accretion rate
distribution, which is heavily dominated by the low-$v$ tail of the
velocity distribution. What could be affected more significantly is
the local density $n^0_{\rm NS}$ of stars, although this too is mostly
determined by the slowest local stars.  In fact, since most NSs are
born in the vicinity of the Galactic plane and move away from it, the
slowest stars, having smaller orbits, will pass more frequently
through the detectable volume. A proper computation of the local
density corrected for the fraction of escaped stars, would require a
knowledge of the birth rate distribution during the lifetime of the
Galaxy (as the oldest stars are more likely to have escaped than the
younger ones). However, it is possible to place an upper limit to the
error made in the local number density by neglecting the fraction of
escaped NSs. This can be done by assuming that all NSs were born at
$t=0$ (Galaxy birth time), and therefore all the ones with velocity
larger than the escape velocity have abandoned the Galaxy by the
present time. Using again equation (\ref{eq:dist}) with a cutoff in
the initial velocity distribution for $v>v_{\rm esc}$, and for an
escape velocity $\sim 500$ km/s (Cordes \& Chernoff 1998), we find
that the correction to the local number density is within 10\%. This
upper limit is well within the uncertainties due to other effects, and
in particular the details of the low-$v$ tail of the velocity
distribution which, according to the estimates made by other groups
(e.g. Arzoumanian et al.  2002) can be more substantial\footnote{The
Arzoumanian et al. distribution has also a more substantial high
velocity tail, and they find that about 50\% of all pulsars will
escape the Galaxy.}.

For the total number $N_{\rm tot}$ of neutron stars populating the
Galaxy at the present time, we use the estimate $N_{\rm tot}\sim 10^9$
based on nucleosynthesis constraints (Arnett et al. 1989), and in
particular on the condition that the number of Type II SNe in the past
be sufficient to account for the total mass fraction of Oxygen
observed in the Galactic disk today. Note that this implies that
either the Type II SN rate was much higher in the past than it is
today\footnote{The currently observed Type II supernova rate from
external Sb and Sc galaxies is about one every 50 years
(e.g. Vandenberg, McClure \& Evans 1987).}, or that the observed rate
today is only a small fraction of the actual core collapse
events\footnote{Arnett et al.  argue that this could be the case if
many core collapse events were no more luminous than SN 1987A, or if a
large fraction of them occurs in dense, obscuring regions, as in the
case of Cas A.}.

\subsection{Emission properties of accreting NSs}

Previous investigations of the expected source counts from accreting
isolated NSs have assumed a blackbody spectrum, although it has been
recognized, since the early work of Zeldovich \& Shakura (1969), that
the spectrum of accreting NSs will depart from a purely Planckian
shape.  Zampieri et al. (1995) worked out the case of pure H
atmospheres with negligible fields at low luminosities, and found that
the emerging spectrum is harder than a blackbody at $T_{\rm eff}$, and
the hardening increases as the temperature decreases. The spectral
shape, as well as this behaviour with $T_{\rm eff}$, is very similar to
that found in unmagnetized, H atmospheres around cooling NSs
(e.g. Romani 1987; Zavlin et al. 1996), as the hardening is not related to the heat source
but to the frequency dependence of the free-free opacity.
The detailed spectra resulting from accretion at low rates onto a
strongly magnetized NS have been computed by Zane et al. (2000). 
They found that the hard tail present in nonmagnetic models with
comparable luminosity is suppressed, and the X ray spectrum is
much closer to a blackbody. 

For a given accretion luminosity, the geometry of the incoming flow is
strongly dependent on the strength of the magnetic field of the star.
In the simplest model (Davidson \& Ostriker 1973), one can assume that
the radius of the accretion column is delineated by the field lines
that pass through the Alfven radius. This leads to an area for the
accretion region $\sim 10^8 B_{12}^{-4/7}\dot{M}_{10}^{2/7}$ cm$^2$, although
it does not account for the fact that the Rayleigh-Taylor instability
can lead to penetration of the accretion material through the field
lines, resulting in much larger emitting areas (Arons \& Lea 1980).
For definiteness, in our calculations we assume a polar cap of area
$\sim 1$ km$^2$.

Given the considerations above, for the purpose of our analysis we use
spectra of unmagnetized, hydrogen atmospheres when considering a model
in which the emission occurs isotropically over the whole star (as
would be typical of low magnetic fields), and blackbody spectra in a
model where accretion occurs through the polar caps (at high magnetic
fields).  Note that, if accretion occurs through a fraction $f$ of the
area of the star, then the corresponding temperature at the star
surface will be larger by a factor $f^{-1/4}$.  For completeness, we
also consider another spectral model which is a combination of
isotropic emission and a power-law. This is empirically motivated, as
many accreting systems display evidence for both a power law and a
thermal component in their spectra (White, Stella \& Parmar 1988),
although note that the formation of standing shocks during the process
of accretion can also give rise to significant power law tails
(Zeldovich \& Shakura 1969).  For definiteness and in agreement with
observations (Campana \& Stella 2000), when considering this model, we
assume that about half of the accretion energy gets reprocessed into a
thermal component, and half comes out in a non-thermal, powerlaw
component.  For the latter, we adopt a photon index of 1.5, up to an
energy of $\sim 10$ keV.

\subsection{Accretion rates}

The standard Bondi formula for the accretion rate onto the surface of
a compact star of mass $M$ is given by the expression 
\beq
\dot{M}_{\rm Bondi}\approx\frac{4\pi G^2 M^2 \rho}{(v^2+c_s^2)^{3/2}}\;,
\label{eq:bondi}
\eeq where $c_s$ and $\rho$ are the sound speed and density of the
ambient ISM, and $v$ is the velocity of the star with respect to the
medium.  Previous investigations of the statistics of accreting NSs
have been based on the Bondi formula to compute accretion
rates\footnote{See however Rutledge (2001) for a discussion of a
subpopulation of highly magnetic $(B\la 10^{14} G)$, accreting NSs
whose accretion rate scales as $v^{1/3}$ rater than $v^{-3}$ as in
Bondi.}.  However, as discussed in \S1, a measurement of the gas
density in the vicinity of the black hole in Sgr A$^*$ indicates that
the Bondi expression highly overestimates the true accretion rate in
that object.

The observational result on Sgr A$^*$ is supported by
three-dimensional MHD simulations by Igumenshchev \& Narayan (2002)
and Proga \& Begelman (2003) which show that, in the presence of a
weak, dynamically unimportant, large-scale magnetic field in the
external medium\footnote{In their simulations, Igumenshchev \& Narayan
assumed that the initial interstellar field had magnetic pressure on the order of 1
percent of the gas pressure. If the field is
stronger, then the effects are only more dramatic.}, the structure of a
spherically-symmetric radiatively inefficient accretion flow is
dramatically different from the Bondi prediction.  As the gas flows
in, the magnetic field is amplified.  When the field reaches
equipartition, it begins to reconnect, heating the gas, and setting up
turbulent motions.  These motions, plus the tension in the amplified
field, strongly suppress the accretion rate onto the black hole.  The
resulting accretion rate scales roughly as \beq \dot{M}\sim \left(
\frac{R_{\rm in}}{R_a}\right)^p \,\dot{M}_{\rm Bondi}\;,
\label{eq:ignar}
\eeq where $R_a\sim GM/c^2_s$ is the ``accretion radius'' at which the
ambient material begins to feel the gravitational influence of the
accreting star, $R_{\rm in}$ is an effective inner radius which is of
order a few to few tens of Schwarzschild radii, and the index $p$ is
uncertain but expected to be less than or of order unity.

The work described above refers to the infall of non-rotating or
slowly-rotating magnetized gas, and is most relevant for the problem
at hand.  However, the main result from the simulations, namely that
the accretion rate is strongly suppressed, is more general and has
been noted also in many studies of strongly rotating flows.  The
evidence has come from a variety of hydrodynamic simulations ---
Stone, Pringle \& Begelman (1999; two-dimensional simulations with
viscosity), Igumenshchev \& Abramowicz (2000; 2D simulations with
$\alpha$ viscosity), Igumenshchev et al.  (2000, 3D $\alpha$
viscosity), Proga \& Begelman (2002; 2D inviscid) --- and MHD
simulations --- Stone \& Pringle (2001; 2D simulations), Machida,
Matsumoto \& Mineshige (2001; 2D and 3D), Hawley, Balbus \& Stone
(2001; 3D), Igumenshchev, Narayan \& Abramowicz (2003; 3D).  All of
these studies indicate a reduced $\dot M$, as in equation
(\ref{eq:ignar}), with $p$ in the range $0.5-1$.

Two different explanations have been advanced for the reduction in
$\dot M$ in radiatively inefficient accretion flows: (i) winds and
outflows (Blandford \& Begelman 1999; Narayan \& Yi 1994, 1995), (ii)
convection (Narayan et al. 2000; Quataert \& Gruzinov 2000).  In the
former explanation, gas accretes at the Bondi rate at the accretion
radius $R_a$.  However, as the gas flows in towards smaller radii, much of
it is lost to an outflow so that only a tiny fraction finally reaches
the central star.  In the latter explanation, convection sets up a
barrier to accretion, such that the net accretion rate is reduced at
all radii out to $R_a$.  Note that, within the latter scenario, a
stationary solution can be obtained as convection changes the
conditions at $R_a$ by increasing the pressure and preventing the gas
from falling in supersonically. The physical picture is somewhat
similar to that envisaged by Ostriker et al. (1976), where preheating
of the infalling gas by the emergent X rays can suppress accretion
(see in this regard also Blaes, Madau \& Warren 1995).  Detailed MHD
simulations provide support for both the outflow picture (Stone \&
Pringle 2001; Hawley et al. 2001) and the convection idea (Machida et
al. 20001; Igumenshchev et al. 2003).  It is possible that both
effects operate, e.g., with convection driving an outflow.

All the simulations described above refer to accretion onto a black
hole.  Most of the results are expected to carry over to the NS case.
If the star is unmagnetized, the estimate of $\dot M$ given in
equation (3) should be valid with essentially no change.  If the star
is strongly magnetized, then equation (3) will only be valid under the
conditions for which the star is able to accrete at a steady pace.
Indeed, in order for a rotating, magnetic star to be able to accrete
steadily, three conditions need to be satisfied (BM, Popov et
al. 2000, Treves et al. 2000, Ikhsanov 2003).  Firstly, the Alfven
radius $r_A$ (where the magnetic energy density is comparable to the
kinetic energy of the accreting gas) must not exceed the accretion
radius $R_a$, otherwise magnetic pressure would dominate everywhere
over the gravitational pull, and the star would be in the so-called
{\em georotator} state\footnote{ This regime was also investigated by
Toropina et al. (1991), who called it ``magnetic plow''. Some
accretion may occur in this regime due to 3D MHD instabilities (Arons
\& Lea 1976, 1980).} (Liuponov 1992).  This condition is typically
fulfilled for the stars in the low-$v$ tail of the velocity
distribution, which dominate the accretion statistics.  Once the
constraint $r_A < R_a$ is satisfied, there are still two barriers that
need to be overcome for accretion to proceed steadily, that is: (i) at
the accretion radius, the ram pressure of the accreting material must
be larger than the radiation pressure from the pulsar (otherwise the
star would be in the {\em ejector} phase), and (ii) the velocity of
the magnetosphere at the Alfven radius must be smaller than the
Keplerian velocity at that same radius. If this latter condition were
not satisfied, the star would act as a {\em propeller} and prevent
material from penetrating inside $r_A$ (Illiaronov \& Sunyaev 1975).

The relative importance of the various phases discussed above during
the life of the NS, depends on a combination of its
velocity, spin period, magnetic field, and accretion rate. In
particular, the strength and the evolution of the magnetic field in
the star play an especially important role. Livio et al. (1998) and
Colpi et al (1998) first pointed out that significant field decay can
cause the NS to linger in the ejector or propeller state for a time longer
than the lifetime of the Galaxy. The fraction of stars which can be in
the accreting phase at the present time is therefore sensitive to both the
initial value of the magnetic field, and to its decay time. For an
initial field strength on the order of $10^{12}$~G, and an
intermediate value of the decay time $\tau_d\sim 10^9$~yr, they found
that about half of the stars are allowed to accrete at the present
time. As a matter of fact, the question of whether the magnetic field
of the stars decays is, by itself, a very controversial issue, both on
theoretical and on observational grounds.  Theoretical predictions
range from an exponential/power-law decay of the field (e.g. Ostriker
\& Gunn 1969; Goldreich \& Reisenegger 1992; Urpin \& Muslimov 1992)
to little or no decay at all within the lifetime of the Galaxy
(e.g. Romani 1990; Srinivasan et al.  1990). Observationally, low
magnetic fields ($\sim 10^9$ G or so), possibly the result of a decay,
have been inferred in low-mass X-ray binaries and millisecond pulsars,
but no firm conclusions have been established yet for isolated
objects. However, if evolution occurs but still leaves a residual
field $B\sim 10^8-10^9$ G, a substantial fraction of the population
should still be able to accrete at the present time (BM, Colpi et
al. 1998).  Keeping these uncertainties in mind, here we assume
that all the NSs have reached the accretion phase at the present time
and are accreting at the rate given by equation (3). 
As a matter of fact, it is sufficient that this condition holds
only for a substantial fraction of the the low-$v$ tail of the
velocity distribution, which is what dominates the accretion statistics.

For the calculations at hand, we need an estimate of the factor by which
the mass accretion rate $\dot M$ onto the NS is suppressed relative to
the Bondi rate. This requires an estimate of $R_{\rm in}$ and $p$
in equation (3).  Abramowicz et al. (2002) found that $R_{\rm in}$ is of
order 50 Schwarzschild radii for a rotating convection-dominated flow;
$R_{\rm in}$ is probably somewhat smaller, by a factor of a few, for a
non-rotating flow.  A more serious uncertainty is in the value of $p$.
The outflow model does not make any specific prediction other than
that $p$ should lie somewhere in the range 0 to 1 (Blandford \&
Begelman 1999).  The idealized convection-dominated accretion flow
model, on the other hand, predicts a precise value: $p=1$ (Narayan et
al. 2000; Quataert \& Gruzinov 2000).  Numerical simulations give a
range of values for $p$, between about 0.5 and 1; the deviation from
unity may be explained within the convection model as the result of
``vertical'' rather than ``radial'' convection (Igumenshchev et
al. 2003).  Finally, we note that a recent analysis of Sgr A$^*$ (Yuan
et al. 2003) indicates that an accretion model with $p\sim0.3$ fits
the observed spectral data and polarization constraints fairly well.

Given the uncertainties, especially in the value of $p$, it is not
possible to derive a precise estimate of $\dot M$ for the accreting
isolated NS problem.  The only statement one can make with confidence,
based on the evidence at hand, is that a radiatively inefficient
accretion flow will {\it not} have $\dot M$ equal to the Bondi rate;
the accretion rate is always suppressed, but it is not clear by precisely how
much.  We thus try a range of suppression factors in this paper: $\dot
M = (10^{-2}, 10^{-3}, 10^{-4}) \dot M_{\rm Bondi}$, with $10^{-3}\dot
M_{\rm Bondi}$ as our best-guess estimate.  The selected range
corresponds roughly to $p\sim0.5-1$.

\section{Source counts: theory confronts observations}

To derive the probability distribution for the accretion rate,
$f(\dot{M},r)$, at distance $r$ from us, we also need an estimate for
the distribution $n_{\rm HI}(r)$ of hydrogen in the ambient medium
surrounding the solar region. Measurements of interstellar absorption
(Paresce 1984; Warwick et al. 1993) suggest that the local
interstellar medium consists of smoothly distributed gas with density
$\sim 0.07$ cm$^{-3}$ up to a distance of about 100 pc, and denser
gas, $\sim 0.5-1$ cm$^{-3}$ beyond that.  As a representative line of
sight to compute source counts per unit solid angle, we will consider
a line of sight with $n_{\rm HI}(r)= 0.07$ cm$^{-3}$ for $r\le 100$ pc
and $n_{\rm HI}(r)=1$ cm$^{-3}$ for $r > 100$ pc. Whereas this is a
good approximation as an average in the local bubble, as
a matter of fact, the ISM is multiphase and comprised of cold
(neutral) gas, as well as warm and hot (ionized) gas.  The hotter
component is more tenuous, and as such the accretion luminosity for
stars embedded within it is further reduced. On the other hand, hot
gas does not contribute much to photoelectric absorption, and the
resulting larger sampling depth along ionized lines of sight tends to
counterbalance the reduction in luminosity.  Here we have assumed for
simplicity all the material to be neutral. A detailed modelling of the
various ISM phases, which would vary depending on the line of sight
under consideration, is beyond the scope of this work.

Let now $C$ be the detector count rate. The number of sources along
a given line of sight which are detectable above the threshold $C$
is 
\beq
\frac{dN}{d\Omega}(>C)=n^0_{\rm NS}\int_0^\infty d\dot{M}
\int_0^{d(\dot{M},C)} dr\,r^2\,f(\dot{M},r)\;,
\label{eq:counts}
\eeq 
where the maximum distance $d(\dot{M},C)$, up to which sources above the
threshold $C$ can be detected, is obtained by solving the implicit equation 
\beq
C=\frac{1}{4\pi d^2(\dot{M},C)}\int_0^\infty L(\nu,\dot{M})A_{\rm
eff}(\nu) e^{-\tau(\nu)}\frac{d\nu}{h\nu}\;. 
\label{eq:dmax}
\eeq
Here $L(\nu,\dot{M})$ is the luminosity of the star, $A_{\rm eff}(\nu)$
is the energy-dependent effective area of the detector under
consideration, and $\tau(\nu)$
is the photoelectric absorption to the source,
 which we compute
using the fit to the cross sections by Morrison \& McCammon (1983). 
For the ROSAT PSPC response, we have taken the version pspcb-gain1-256\footnote{
ftp://legacy.gsfc.nasa.gov/caldb/data/rosat/pspc/cpf/matrices/}
and used it in Eq.(\ref{eq:dmax}) by performing an interpolation to
the data. 

As a first step towards estimating the contribution of the various
effects described above to the source statistics, in Figure 2 we compare
the number counts obtained under the same assumptions made by BM
(i.e. NO pulsar velocity distribution, Bondi accretion rate, and blackbody spectra) with 
those obtained assuming the CC velocity distribution and atmospheric spectra
but still with the Bondi accretion rate. As the figure shows, 
despite the reduction in the number counts due to the higher mean velocity,
the predicted rates are still well above the limits set by the ROSAT
All Sky Survey. In particular, in the case of isotropic emission with
atmospheric spectra, the reduction in the counts due to the higher mean velocities is
considerably offset by the increase due to the atmospheric rather than
BB spectra. In fact, as explained above, for the same value of the star
temperature, the atmospheric spectra are harder than BB, hence
producing more emission in the X-ray band.  

In Figure 3, we show the predicted source counts once the correction
to the Bondi rate discussed in \S 2.3 is introduced.
Note that, while in Figure 2 we showed the counts per unit solid
angle, here we have considered the numbers over the entire sky.  This is
because, in this case, the distances over which the sources are
detectable are comparable to the size of the local bubble, over which
the density distribution can be taken as isotropic, and therefore the
line of sight that we have used can be considered as representative
of the whole sky.  It can be seen from Figure 3, that the predicted
counts are consistent with a null detection at the ROSAT PSPC
sensitivity.

As discussed in \S 2.3, a rough estimate for the reduction in the
Bondi rate is $\sim 10^{-3}$; however, for completeness, we explore
the effects on the source counts for the range of $\dot{M}\sim
10^{-2}-10^{-4} \dot{M}_{\rm Bondi}$.  This is shown in Figure 4, for
the case of polar cap emission and the CC velocity
distribution. Whereas for $\dot{M}\la 10^{-3} \dot{M}_{\rm Bondi}$
the counts are consistent with no detection at the ROSAT sensitivity,
a rate $\dot{M}\sim 10^{-2}\dot{M}_{\rm Bondi}$ would still require
that a good fraction of the sources is not in the accretion phase for
most of the time.

\section{Summary}

We have revisited the problem of the statistics of old and isolated
neutron stars, which ought to be detectable in X-rays as a result of
accretion from the ISM.  We have used the recent estimates for the
velocity distribution made by Cordes \& Chernoff (1998), we have gone
beyond the blackbody assumption for the spectra, using atmospheric
models, and we have used a ``revised Bondi'' accretion rate, according
to the results of MHD simulations and supported by direct observations
in the case of accretion around supermassive black holes.  Whereas the
first two effects together do not result in a significant reduction of
the predicted number counts, the modification to the Bondi formula
dramatically reduces the number of observable sources.

As discussed in \S 2.3, other effects can also
influence the detection statistics. Particularly important is the
role played by the magnetic field of the star. This determines the
spin evolution of the star itself, and the consequent fraction of the
lifetime during which the star can be in a regime of steady accretion,
rather than in other phases such as ejector or propeller.  The
importance of these various phases during the evolution of the NS life
strongly depends on the evolution of the magnetic field within the
star, which is by itself a very controversial issue. The source counts
that we have computed here should therefore be considered as upper
limits in that they are representative of a population that is
accreting for most of its lifetime.

The number of observable sources that we have derived, with the inclusion 
of the revised velocity distribution, atmospheric spectra and reduced
Bondi rate, is consistent with the detection
of only a few candidates in the ROSAT all sky survey.  As of
now, there are only two objects which have been confirmed to
be isolated NSs (see Rutledge et al. (2003) for a comprehensive review of the
observations so far); moreover it is hard to firmly establish that
they are actually powered by accretion and not by cooling.
In the X-ray range, the spectrum from an accreting star looks quite similar
to that of a cooling NS; however, Zane et al. (2000) noted that 
the two scenarios can be distinguished from optical observations, as
accretion spectra have been found to display an excess emission over
the best fitting blackbody X-ray distribution, which, on the other
hand, does not appear in cooling models with similar temperatures.

\acknowledgements We thank Omer Blaes, Bob Rutledge, Roberto Turolla
and Silvia Zane for very useful discussions related to various aspects
of this work. We particularly thank Omer Blaes for providing us with
a routine for the space density distribution. We also thank
the referee for helpful comments on the manuscript.
RP aknowledges support by the Harvard Society of Fellows and a research
grant by the Harvard-Smithsonian Center for Astrophysics.
RN was supported in part by NASA grant NAG5-10780.

\newpage

\clearpage
\begin{figure}[t]
\plotone{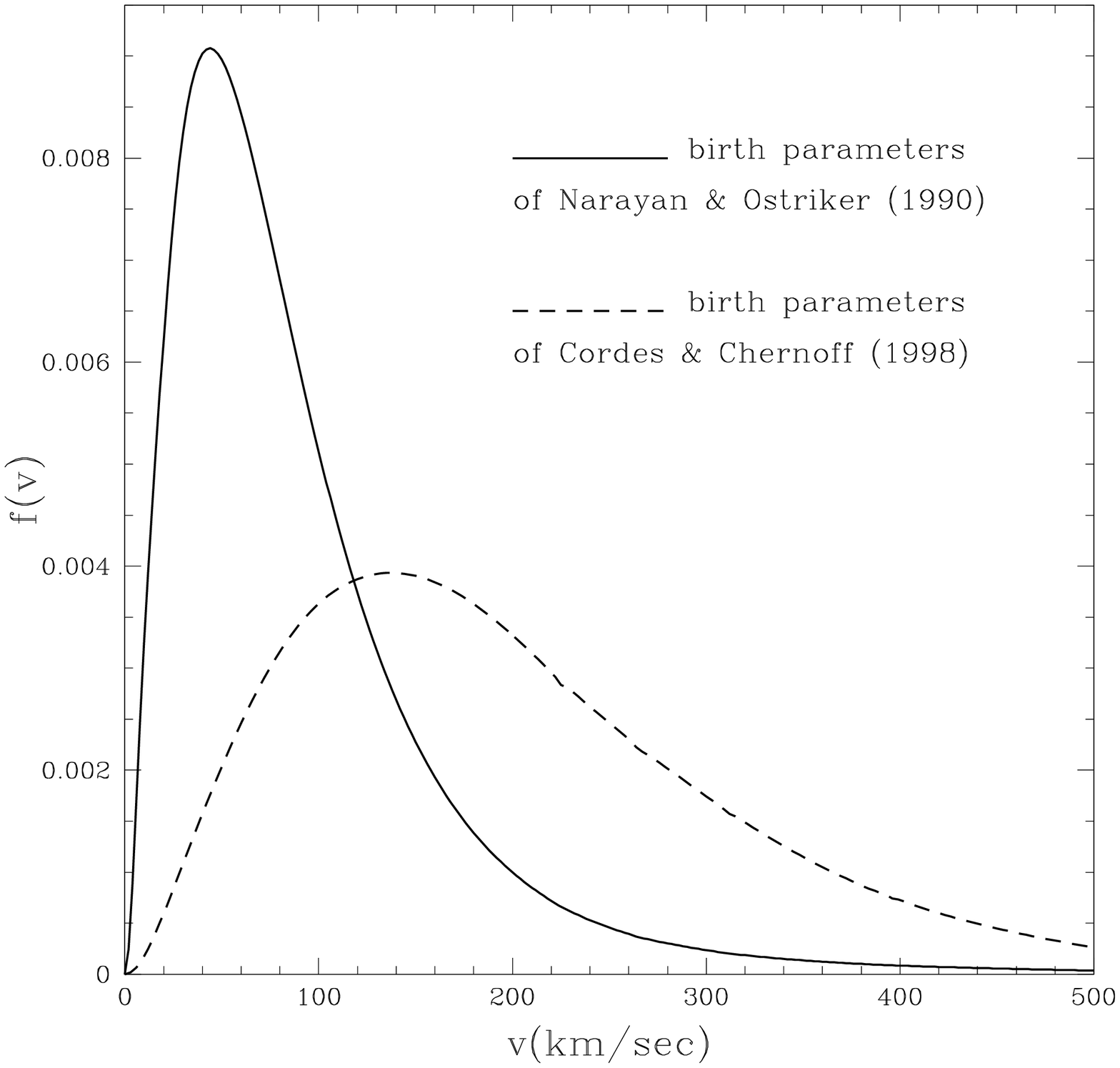}
\caption{Comparison between the speed distribution of NSs in
the solar neighborhood ($R=8.5$ kpc, $z=0$) obtained assuming the
Narayan \& Ostriker (1990) initial distribution and the Cordes \&
Chernoff (1998) distribution.}

\end{figure} 

\begin{figure}[t]
\plotone{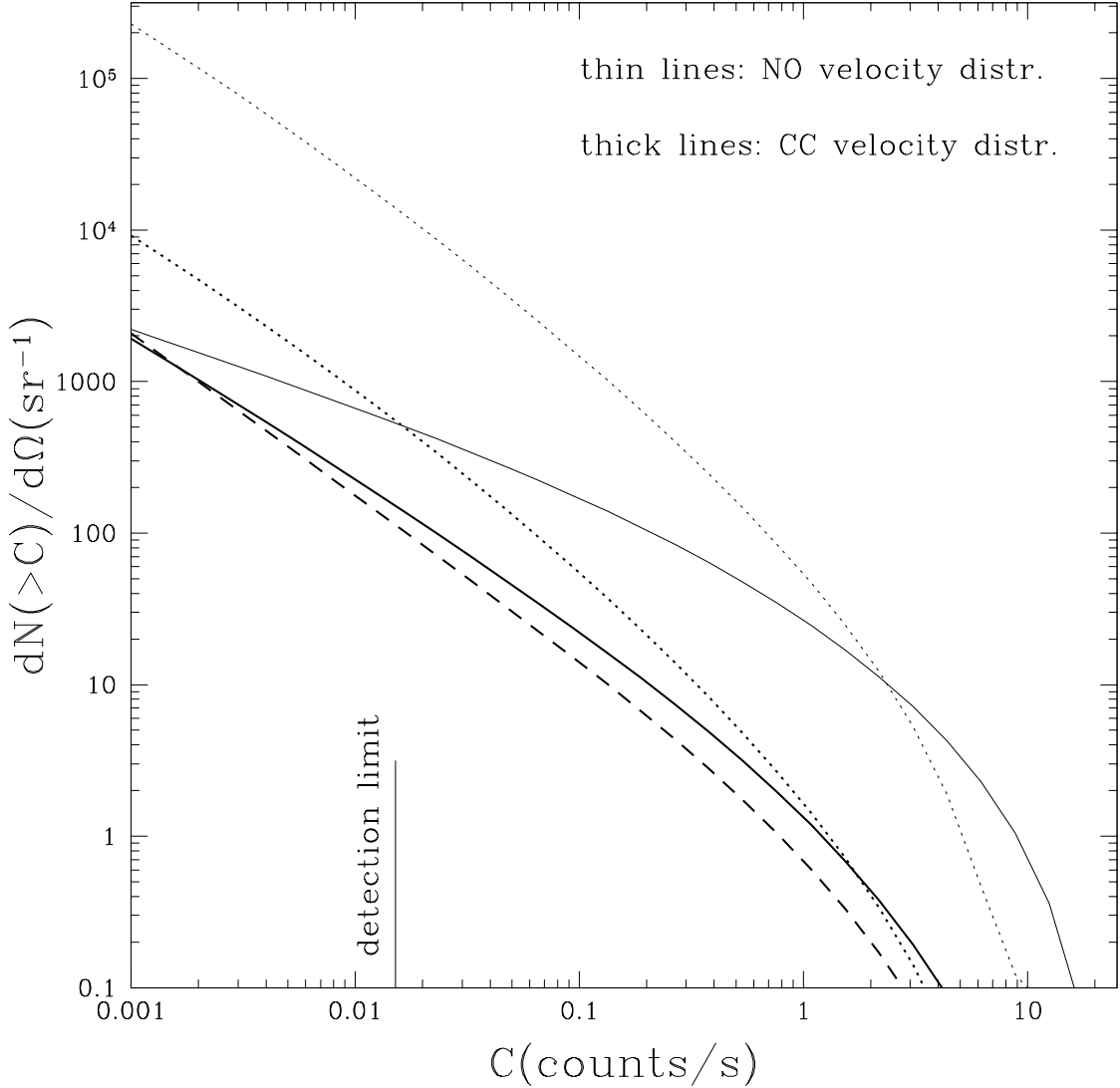}
\caption{The combined effect on source statistics of the
higher mean velocities and various spectral energy distributions, 
shown for the cumulative counts in the 0.1-2.4 keV band of the
ROSAT PSPC detector.  The thin lines assume the NO velocity
distribution and BB spectra both for the case of isotropic emission
(solid line) and polar cap emission (dotted line). This is the case
studied by BM.  The thick lines assume the CC velocity distribution
and consider various possiblities for the spectral energy
distribution: blackbody in the case of polar cap emission (dotted
line), unmagnetized hydrogen atmosphere in the case of isotropic emission (solid
line), and a combination of isotropic, atmospheric emission with a
power law (dashed line).  In all cases, the accretion rate is assumed
to be equal to the Bondi rate.  In the case of isotropic emission, for
count rates $\la 0.001$ s$^{-1}$, the reduction in the counts due to
the increased mean velocity is compensated by the increase due to the
harder atmospheric spectrum.}
\end{figure} 

\begin{figure}[t]
\plotone{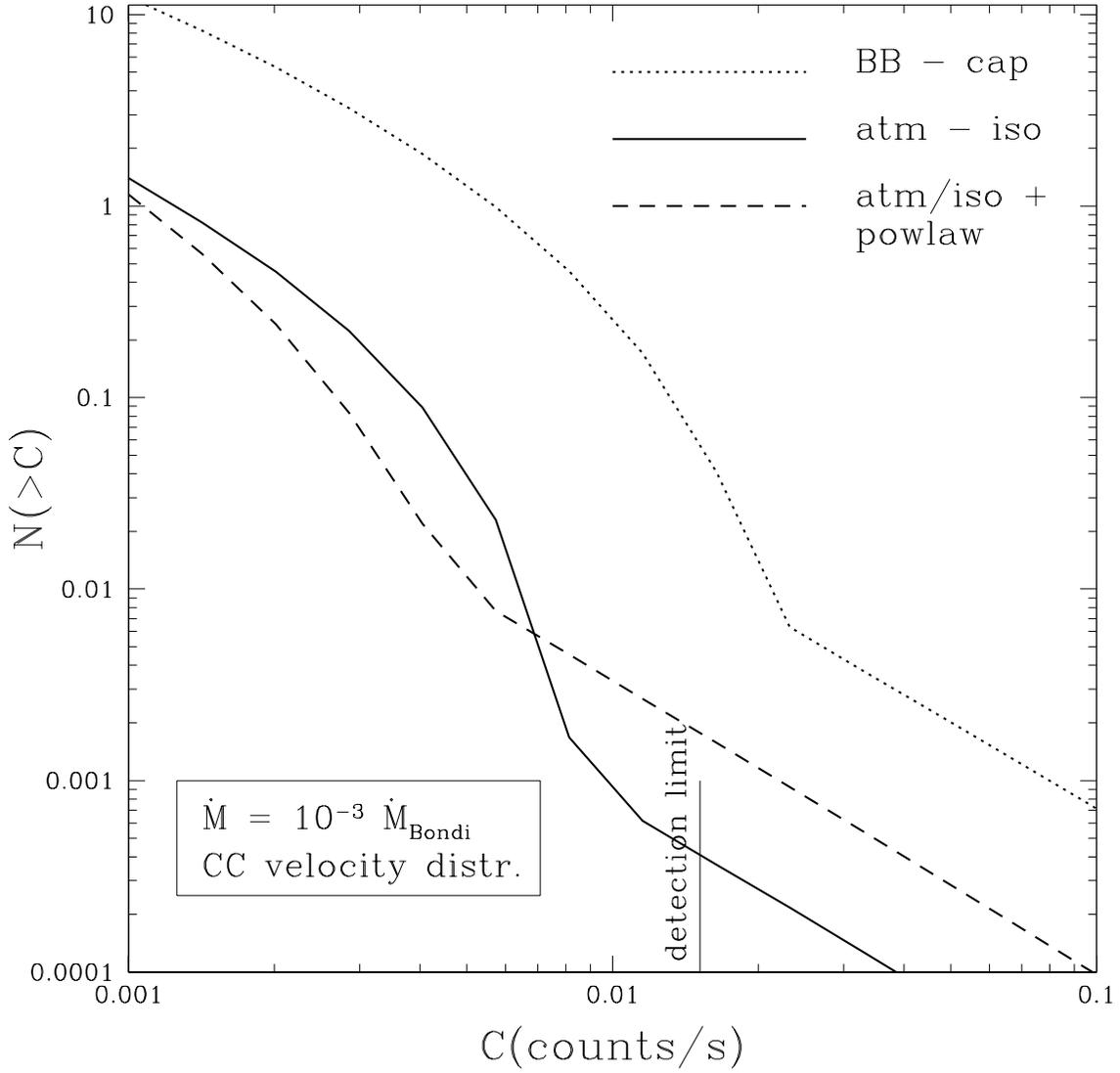}
\caption{Cumulative source counts in the 0.1-2.4 keV band of the
ROSAT PSPC detector over the entire sky (assuming that the line of
sight that we have considered is representative of the whole sky over
a small distance scale). The computation of these counts includes the
CC velocity distribution, an atmospheric spectrum for the isotropic
component of the emission, and a mass accretion rate 
reduced below the Bondi rate by a factor of
$10^{-3}$, according to the discussion in \S2.3.}
\end{figure} 

\begin{figure}[t]
\plotone{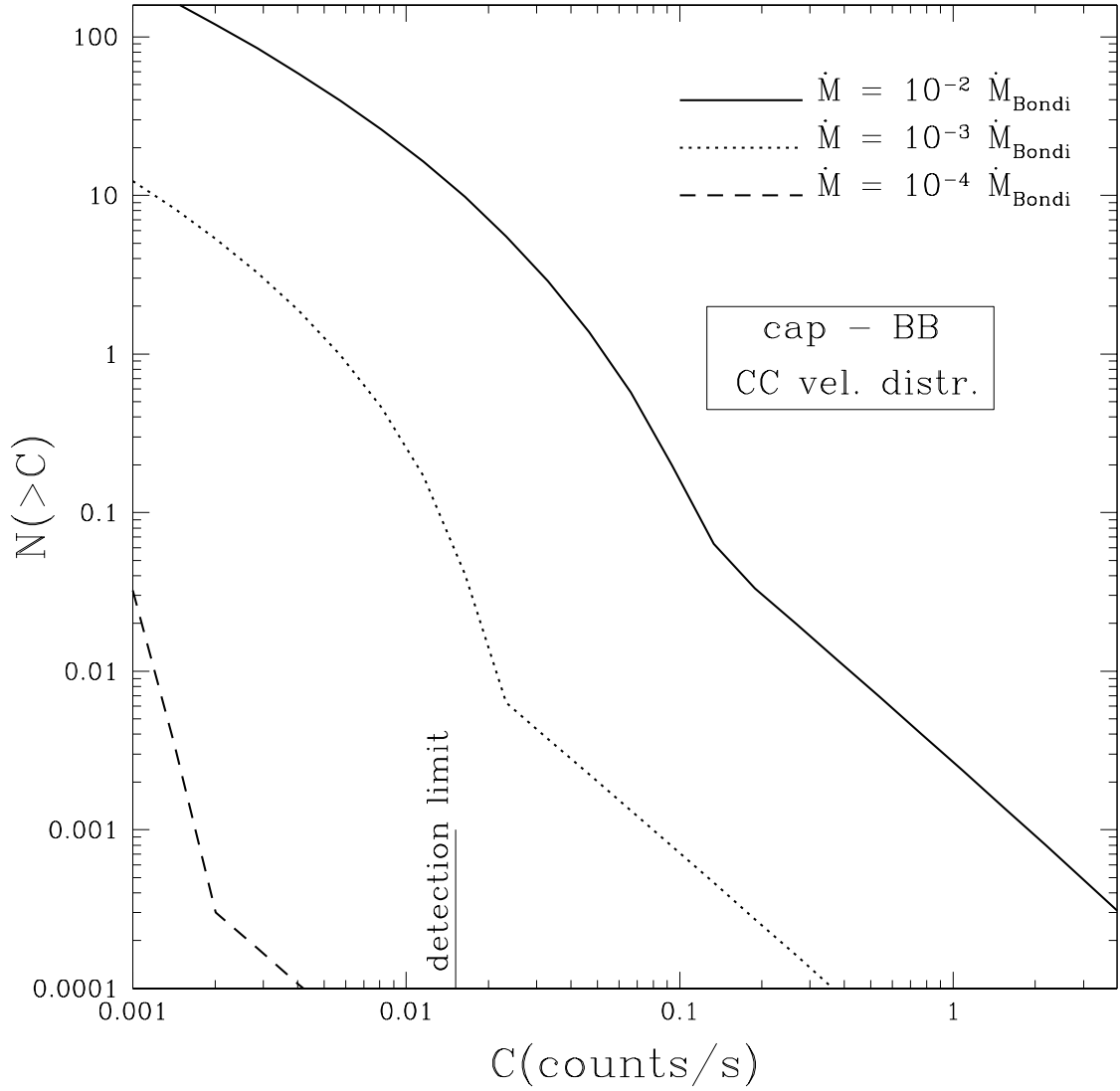}
\caption{The effect of a suppressed mass accretion rate on the 
number counts in the 0.1-2.4 keV band of the ROSAT PSPC detector, 
shown for the case of polar cap emission with a blackbody spectral energy
distribution, and for the CC velocity distribution.}
\end{figure}

\end{document}